\documentclass[reprint,amsmath,amssymb,
 aps,nofootinbib,prd ]{revtex4-2}
\usepackage{amsmath,amscd, empheq}
\usepackage{amsfonts}
\usepackage{amssymb}
\usepackage{graphicx}
\usepackage{color}
\usepackage[normalem]{ulem}
\usepackage{hyperref}
\usepackage{enumerate}
\usepackage{mathtools}
\usepackage{dcolumn}
\usepackage{bm}
\usepackage{shuffle}

\begin{document}

\title{Color-kinematics and double-copy relations for selfdual solutions}

\author{Joris Raeymaekers}

\affiliation{CEICO, Institute of Physics of the Czech Academy of Sciences,\\  Na Slovance 2, 182 00 Prague 8, Czech Republic}

\email{joris@fzu.cz}


\begin{abstract}
We clarify the relation between the classical double copy and  the double copy  for amplitudes in the setting of selfdual gauge and  gravity theories. 
To this end we construct explicit all-order perturbative solutions in these theories and  show that they are  related by a version of color-kinematics duality. This relation can be expressed in a double-copy form and  embodies the most general manifestation of the   selfdual classical double copy. Our classical double copy relations directly  lead to known amplitude double-copy relations, both for Berends-Giele currents and on-shell amplitudes, through the perturbiner expansion.
\end{abstract}

\maketitle

\section{Introduction}
The discovery of the Kawai-Lewellen-Tye-like  double copy relations connecting amplitudes in gauge and gravity  theories, and the underlying  color-kinematics duality, has revolutionized our understanding of  the on-shell structure of these theories (see \cite{Bern:2019prr} for a review and list of  references). In a subsequent development,  it was realized first in \cite{Monteiro:2014cda} that classical solutions of gauge and gravity theories are in some cases related by  classical double copy relations. We refer to \cite{White:2024pve} for a review and a guide to the extensive literature.
While this classical double copy  bears a formal similarity  to the double-copy relations for amplitudes, the direct relation between the two  is not well-understood in general, though  it has been clarified in the case of  radiation fields of point particles \cite{Luna:2016due}, see also \cite{Goldberger:2016iau,Bautista:2019tdr,Shen:2018ebu,PV:2019uuv}.  

A general and   direct relation between classical solutions and tree-level amplitudes\footnote{Perturbiner methods have recently been extended to  loop amplitudes \cite{Lee:2022aiu,Gomez:2022dzk,Gomez:2024xec}.} exists in the form of  the perturbiner expansion \cite{Rosly:1996cp,Rosly:1996vr,Selivanov:1997aq,Rosly:1997ap}, reviewed in  \cite{Mizera:2018jbh}.  More precisely, this expansion extracts from perturbative solutions to the equations of motion  the Berends-Giele (BG) currents \cite{Berends:1987me},  which are amplitudes where one leg  is off-shell. As a next step, on-shell amplitudes can be easily computed from the BG currents.

Motivated by these considerations, we aim in this work to use perturbiner methods to clarify the relation between classical and amplitude double copy relations. We focus on the setting of selfdual Yang-Mills and gravity theories, which has  the advantage that both theories, while nonlinear, are also integrable and therefore expected to be fully tractable.    As stressed in \cite{Luna:2016hge}, in the nonlinear setting  the classical double copy should be viewed as  a  relation between the general perturbative structure of  
both theories, rather than a one-to-one map between solutions. In the selfdual sector of interest, an important step towards realizing this idea was taken in \cite{Monteiro:2011pc}, where the kinematic algebra which underlies color-kinematics duality for selfdual theories was identified. Color-kinematics duality for selfdual theories was analyzed at the level of actions in \cite{Borsten:2023paw}.   In this work we will build on \cite{Monteiro:2011pc} and,  using recently developed techniques for solving BG recursion relations \cite{Frost:2020eoa},  derive a novel all-order perturbative solution to Plebanski's equation \cite{Plebanski:1975wn} for selfdual gravity. This solution   can be   written as  a manifest double-copy of the Bardeen-Cangemi \cite{Bardeen:1995gk, Cangemi:1996rx} perturbative Yang-Mills solution.  Our result implies\footnote{The converse is not necessarily true, i.e. knowledge of the BG currents does not fully determine the all-order classical solution in general.}, through the use of the perturbiner expansion, known double copy relations between BG currents  \cite{Cho:2021nim,Correa:2024mub} and on-shell amplitudes in both theories.  
 
\section{Selfdual Yang-Mills and gravity}
In this section we briefly review the selfdual Yang-Mills and Einstein equations,  referring to \cite{Krasnov:2016emc} for more details. We will work in mostly minus signature and make use of bispinor notation\footnote{Spinor indices are raised and lowered according to
\begin{equation}
    \mathfrak{p}_\alpha = \mathfrak{p}^\beta \epsilon_{\beta\alpha}, \qquad  \mathfrak{p}^\alpha = \epsilon^{\alpha\beta} \mathfrak{p}_\beta,\qquad  \epsilon_{12} = \epsilon^{12} = 1.\nonumber
\end{equation} and similarly for dotted indices.} with
\begin{equation}
\sigma^\mu_{\alpha\dot\alpha}= {1\over \sqrt{2}} (1, \sigma_i )    , \ \ \  x^{\alpha\dot \alpha} = x^\mu \sigma_\mu^{\alpha\dot \alpha} \equiv  \left(\begin{array}{cc} u & w\\\bar w& v \end{array}\right),\label{coorddef}
\end{equation}
in terms of which the Minkowski metric reads
\begin{equation}
    ds^2_M = \epsilon_{\alpha\beta}\epsilon_{\dot \alpha\dot \beta} d x^{\alpha\dot \alpha} d x^{\beta\dot \beta} = 2 (d u dv - d w d \bar w).
\end{equation}
We also use the spinor helicity formalism, where a   massless on-shell momentum $p$ with  $p^2 =0$  is decomposed as a product of spinors and where  angle- and square-bracket inner products are introduced as
\begin{equation}
p_{\alpha\dot \alpha}= \mathfrak{p}_\alpha\tilde{\mathfrak{p}}_{\dot \alpha},\ \ \ [12]= \mathfrak{p}_1^\alpha \mathfrak{p}_{2\alpha},\ \ \ \langle 12 \rangle  = \tilde{\mathfrak{p}}_1^{\dot \alpha} \tilde{\mathfrak{p}}_{2\dot\alpha}.
\end{equation}
Polarization tensors for gluons and gravitons 
 depend on an arbitrary massless reference momentum, reflecting gauge invariance. In what follows we will consider all polarizations tensors  to involve  the same reference momentum $q$,   which we fix once and for all and decompose into spinors $\mathfrak{q}_\alpha, \tilde{\mathfrak{q}}_{\dot \alpha}$ as
\begin{equation}
q_{\alpha\dot\alpha} =  \mathfrak{q}_\alpha\tilde{\mathfrak{q}}_{\dot \alpha}.
\end{equation}
In terms of this reference momentum,  the polarization tensors for a gluon resp. graviton with momentum $p\neq q$ are given by
\begin{align}
      e_+ ( p)_{\alpha\dot\alpha}= &{\mathfrak{q}_\alpha \tilde {\mathfrak{p}}_{\dot \alpha} \over [ q p ]} , &  e_{++} (p)_{\alpha\dot\alpha \beta \dot \beta} =&{\mathfrak{q}_\alpha  \mathfrak{q}_\beta \tilde {\mathfrak{p}}_{\dot \alpha} \tilde {\mathfrak{p}}_{\dot \beta} \over [ q p ]^2}\nonumber\\
            e_- ( p)_{\alpha\dot\alpha}= &{{\mathfrak{p}}_{\alpha} \tilde{\mathfrak{q}}_{\dot \alpha}   \over \langle q p\rangle } ,&   e_{--} (p)_{\alpha\dot\alpha \beta \dot \beta} =&{\mathfrak{p}_\alpha  \mathfrak{p}_\beta \tilde {\mathfrak{q}}_{\dot \alpha} \tilde {\mathfrak{q}}_{\dot \beta} \over \langle q p \rangle^2}.\label{pols}
      \end{align}

We will now write the equations of motion for (anti-) selfdual Yang-Mills and gravity fields in a gauge which depends on the chosen reference spinor $\mathfrak{q}_\alpha$ (resp. $\tilde{\mathfrak{q}}_{\dot \alpha}$).  We will restrict our attention to the selfdual sector, the anti-selfdual one being analogous.  While it is of course possible to Lorentz rotate and rescale the reference spinor to a convenient value such as  $\mathfrak{q}^\alpha =\delta^\alpha_1$,  we will keep it  general for easy comparison with the amplitude literature. We first introduce a spinor $\mathfrak{r}^\alpha$ so that $(\mathfrak{q}^\alpha , \mathfrak{r}^\alpha)$ form an orthonormal frame in the sense that
\begin{equation}
    \mathfrak{q}^\alpha  \mathfrak{r}^\beta- \mathfrak{q}^\beta \mathfrak{r}^\alpha = -\epsilon^{\alpha\beta}.
\end{equation}
We then have
\begin{equation}
     \mathfrak{q}^\alpha \partial_{\alpha\dot \alpha} = {\partial \over \partial x^{\dot \alpha}} \ {\rm with\ } x^{\dot \alpha} := \mathfrak{r}_\alpha x^{\alpha \dot\alpha}.
\end{equation}
We make the following ans\"atze for the Yang-Mills and gravitational fields  in terms of `prepotential' functions $\Phi$ and $\Psi$:
\begin{eqnarray}
    A_{\alpha \dot\alpha} &=& g  \mathfrak{q}_\alpha \partial_{\dot \alpha} \Phi \label{AitoF}\\ 
   g_{\alpha\dot\alpha \beta\dot \beta}&=&  \epsilon_{\alpha\beta}\epsilon_{\dot \alpha\dot \beta}  +   \kappa \mathfrak{q}_\alpha \mathfrak{q}_\alpha \partial_{\dot \alpha} \partial_{\dot \beta}   \Psi.\label{hitoPsi}
\end{eqnarray}
These  ans\"atze satisfy the Lorentz resp. transverse-traceless gauge conditions and solve the linearized selfduality conditions
$ \epsilon^{\dot \alpha\dot \beta } \partial_{[\alpha\dot \alpha} A_{\beta \dot \beta]}=   \epsilon^{\dot \alpha \dot \beta } \partial_{[\alpha\dot \alpha} \Gamma_{\beta \dot \beta] \nu }^\mu=0$, provided that the prepotentials satisfy d'Alembert's equation $\partial^2\Phi=\partial^2\Psi=0$, with $\partial^2 \equiv\partial^{\alpha \dot \alpha} \partial_{\alpha \dot \alpha}$. Taking  $\Phi, \Psi$ to be  appropriately normalized plane waves  $\Phi = J e^{i p \cdot x}, \Psi = \mathcal{J} e^{i p \cdot x}$ leads to  positive helicity gluon resp. graviton fields, i.e.
\begin{eqnarray}
    J =  - {i\over [ qp]^2} &\Rightarrow& 
A_{\alpha \dot\alpha} =g  e_+ (q; p)_{\alpha\dot\alpha} e^{i p \cdot x} \nonumber \\
  \mathcal{J}  = - {1\over [ qp]^4} &\Rightarrow& 
h_{\alpha \dot\alpha \beta\dot\beta} =\kappa e_{++} (q; p)  e^{i p \cdot x} .\label{Jvals}
\end{eqnarray}
The full nonlinear selfdual equations reduce to
\begin{eqnarray}
    \partial^2\Phi &=& g \epsilon^{\dot \alpha \dot \beta}\left[ \partial_{\dot \alpha} \Phi,  \partial_{\dot \beta}\Phi\right]\label{SDYM}\\
 \partial^2\Psi &=& {\kappa\over 2}  \epsilon^{\dot \alpha \dot \beta} \left\{ \partial_{\dot \alpha} \Psi,  \partial_{\dot \beta} \Psi \right\},\label{SDgrav}
\end{eqnarray} 
where, in the second line, we have introduced a Poisson bracket  as 
\begin{equation}
     \{ F, G \} =   \epsilon^{\dot \alpha \dot \beta}   \partial_{ \dot \alpha} F \partial_{ \dot \beta} G.\label{PB}
\end{equation}
The equations (\ref{SDYM},\ref{SDgrav}) are equivalent to the standard selfdual Yang-Mills  resp.  gravity equations.  For example, choosing $\mathfrak{q}^\alpha=\delta^\alpha_1$ leads to $x^{\dot 1} =u, x^{\dot 2} = w$ (cfr. (\ref{coorddef})) and the gravity equation  becomes  Plebanski's second heavenly  equation \cite{Plebanski:1975wn},
\begin{eqnarray}
  (\partial_{uv} - \partial_{w\bar w}) \Psi     &=& {\kappa\over 2} \left( \partial_u^2 \Psi \partial_w^2\Psi - (\partial_{uw}   \Psi)^2\right).
\end{eqnarray}
    The equations (\ref{SDYM},\ref{SDgrav})  can be suggestively rewritten in momentum space. To simplify the formulas we will treat the  momentum dependence as a continuous index, writing e.g. $f^k := f(k), \delta^k_l := \delta (k-l) $ and adopting an Einstein summation convention 
    \begin{equation}
        f^k g_k = \int d^4 k f(k) g(k).
    \end{equation} Eqs.  (\ref{SDYM},\ref{SDgrav}) then take the form
    \begin{eqnarray}
      \Phi^{k} &=&2  g {  F^k_{k_1k_2} \over s_{12}}  \Phi^{k_1}  \Phi^{k_2}\label{momspcs} \\
        \Phi^{ak} &=& g {f^a_{a_1a_2} F^k_{k_1k_2} \over s_{12}}  \Phi^{a_1k_1}  \Phi^{a_2k_2}\label{momspcd} \\
        \Psi^{k} &=& \kappa {\tilde F_{k_1k_2} F^k_{k_1k_2} \over 2 s_{12}}  \Psi^{k_1}  \Psi^{k_2} ,\label{SDmom}
    \end{eqnarray}
where $s_{ij}:=(k_i + k_j)^2$.
The first two equations are equivalent, but it will be convenient in what follows to separately consider the equations   for the Lie-algebra valued scalar $\Phi^k = \Phi^{ak} T_a$  and for it's color-components $\Phi^{ak}$. The $f^a_{a_1 a_2}$ are the structure constants of the Yang-Mills gauge group, while the $F^k_{k_1 k_2}$ and $\tilde F_{k_1 k_2}$ arise from the Poisson bracket of plane waves:
\begin{equation}
     \{ e^{i k_1\cdot  x}, e^{i k_1 \cdot x} \} = \sum_k   F^k_{k_1 k_2}   e^{i k \cdot x}.\label{PBexp}
\end{equation}
In particular,
\begin{equation}
    F^k_{k_1 k_2} = \tilde F_{k_1 k_2}  \delta ^k_{k_1+k_2},  \ \ \ \tilde F_{k_1 k_2}=\mathfrak{q}^\alpha\mathfrak{q}^\beta \epsilon^{ \dot \alpha \dot \beta} k_{1 \alpha \dot \alpha} k_{2 \beta \dot \beta}.\label{Fdef}
\end{equation}
 For on-shell momenta, $k_1^2 = k_2^2=0$, the $\tilde F_{k_1k_2}$ reduce to
  \begin{equation}
    \tilde F_{k_1k_2}=  - [q1][q2]\langle 1 2\rangle\label{XitoS}.
 \end{equation}
 The $  F^k_{k_1 k_2} $ are in fact structure constants of an infinite-dimensional `kinematic' Lie algebra \cite{Monteiro:2011pc}, namely the infinite-dimensional  algebra $sdiff_2$  of infinitesimal area-preserving diffeomorphisms. To see this, 
 we consider the symplectic form corresponding to (\ref{PB})
 \begin{equation}
     \Omega =-{1\over 2} \epsilon_{\dot \alpha \dot\beta} dx^{\dot\alpha}\wedge  dx^{\dot\beta}.
 \end{equation}
Any function $F(x^{\dot \alpha})$ gives rise to a Hamiltonian vector  field  $L_F $ through $i_{L_F} \Omega = d F,$ which preserves the area form $\Omega$ in the sense that $\mathcal{L}_{L_F} \Omega=0.$ The Lie bracket of the associated vector fields is
\begin{equation}
     [ L_F, L_G]= - L_{\{F,G\} }.
\end{equation}
Therefore the $F^k_{k_1 k_2}$ are structure constants of $sdiff_2$ in a basis associated to exponential functions. In another basis, associated to monomials, the structure constants reduce to those of the algebra $w_{1+\infty}$ \cite{Bakas:1989xu}. Indeed, defining
\begin{equation} 
 w^{(i)}_m = (x^{\dot 1})^{m+i-1} (x^{\dot 2})^{i-1}
 \end{equation}
 we have 
 \begin{equation} 
 \{  w^{(i)}_m,  w^{(j)}_n \} = \left((j-1)m -(i-1)n\right) w^{(i+j-2)}_{m+n}.\label{Winftgenrs}
 \end{equation} 
The interpretation of $F^k_{k_1 k_2}$ as kinematic structure constants makes it clear that the selfdual Yang-Mills and Einstein  equations in the form (\ref{momspcd},\ref{SDmom}) are related by a type of color-kinematics duality.

\section{All-order perturbative solutions}
In this section we construct perturbative solutions to the selfdual equations (\ref{momspcs}-\ref{SDmom}). We expand the prepotentials as
\begin{equation}
     \Phi = \sum_{n=1}^\infty g^{n-1} \Phi_{(n)} , \qquad \Psi =  \sum_{n=1}^\infty \left({\kappa\over 2}\right)^{n-1} \Psi_{(n)}.
\end{equation}
We write the $n$-th order solution as an integral over momenta of  a kernel as follows
\begin{eqnarray}
    \Phi_{(n)}^k &=& K^{ k}_{k_1 \ldots k_n}  \Phi_{(1)}^{k_1} \ldots  \Phi_{(n)}^{k_n}\label{Kcsdef}\\
      \Phi_{(n)}^{ak} &=&  {1\over n!}  K^{ a k}_{a_1 \ldots a_n,k_1 \ldots k_n} \Phi_{(1)}^{a_1 k_1} \ldots  \Phi_{(n)}^{a_n k_n}\label{Kcddef}\\
      \Psi_{(n)}^k &=& {1\over n!} \mathcal{K}^{ k}_{k_1 \ldots k_n} \Psi_{(1)}^{k_1} \ldots  \Psi_{(n)}^{k_n} \label{Kgravdef}
\end{eqnarray}
In the Yang-Mills case we have defined both a color-stripped kernel $K^{k}_{k_1 \ldots  k_n}$ and a color-dressed one $K^{ a k}_{a_1 \ldots a_n, k_1 \ldots  k_n} $, and for later convenience we have included combinatorial factors in the last two definitions. To further simplify the formulas we will use the shorthand notation
\begin{eqnarray}
   K^k  (1 \ldots n) &:=& K^{ k}_{k_1 \ldots k_n}  \\
    K^{ak} (1\ldots n) &:=& K^{ a k}_{a_1 \ldots a_n,k_1 \ldots k_n}    \\
 \mathcal{K}^k (1 \ldots n) &:=&   \mathcal{K}^{ k}_{k_1 \ldots k_n}. 
\end{eqnarray}
We will also use straightforward generalizations of  these definitions where the argument $1 \ldots n$ is replaced by a general word (i.e. a sequence of positi ve integer letters), and is linearly extended to combinations of words $P,Q$, i.e. $ K^k  (a P + b Q)=a  K^k  ( P ) +b  K^k  ( Q ) $.

The color-stripped kernel $K^k$ must satisfy nontrivial properties which guarantee that the RHS of (\ref{Kcsdef}), which formally belongs to the universal enveloping algebra, in fact reduces to an element of the Lie algebra. These are the shuffle symmetries \cite{Berends:1988zn,Lee:2015upy}
\begin{equation}
    K^k  ( P \shuffle Q)=0,\label{Kcssymm}
\end{equation}
where the shuffle product of two  words $P$ and $Q$ is the sum over all permutations of $P  Q$ which preserve the orderings of both words, for example $12\shuffle 3 = 123 + 132 + 312$.  
Both the color-dressed kernel  $K^{ a k}$ 
and the gravity kernel $\mathcal{K}^{ k}$ 
are  symmetric under permutations of the letters, i.e.
\begin{equation} K^{ak}  (\sigma \cdot P )  =  K^{ak}  (P),   \quad  \mathcal{K}^k  (\sigma \cdot P) = \mathcal{K}^k  (P), \label{Kcdsymm}\end{equation}
where \(\sigma\cdot P \) is an arbitrary permutation of the word $P$.

Substituting (\ref{Kcsdef}-\ref{Kgravdef}) in  the equations (\ref{momspcs}-\ref{SDmom}) leads to quadratic recursion relations for the kernels:
\begin{eqnarray}
       K^{k}  (P) &=&{ 2 \over s_{P}}\sum_{QR=P }  {F^k_{lm}}  K^{ l}  (Q) K^{m} (R)\label{Kcsrec}\\
       K^{ ak}  (\mathcal{P})&=&  {1 \over   s_{\mathcal{P}}}\sum_{\mathcal{Q}\cup \mathcal{R}=\mathcal{P} } {f^a_{bc}F^k_{lm} }  K^{ bl}  (\mathcal{Q}) K^{ cm} (\mathcal{R}) \label{Kcdrec}\nonumber \\ \\
         \mathcal{K}^{ k} (\mathcal{P}) &=& {1 \over   s_{\mathcal{P}}} \sum_{\mathcal{Q}\cup \mathcal{R}=\mathcal{P}} {\tilde F_{lm} F^k_{lm}}    \mathcal{K}^{ l} (\mathcal{Q}) \mathcal{K}^{ m}    (\mathcal{R}) ,\label{Kgravrec}\nonumber \\ 
    \end{eqnarray}
where $s_P = (k_{P_1} + \ldots k_{P_{|P|}})^2$  and $|P|$ denotes the length of the word $P$.  The starting values for the above recursions are 
\begin{equation}
    K^{ k}(1)= \mathcal{K}^{ k}(1)= \delta^k_{k_1},\qquad K^{ak}(1)=\delta^a_{a_1} \delta^k_{k_1}.
\end{equation}
In the above expressions,  Latin letters $P,Q,\ldots$ stand for arbitrary words while calligraphic ones $\mathcal{P},\mathcal{Q}, \ldots$ denote ordered words, whose letters are in increasing  order. The sum over $QR=P$ runs over all deconcatenations of   $P$ into non-empty words, while the sum over $\mathcal{Q}\cup \mathcal{R}= \mathcal{P}$ runs over ways to distribute the letters of  $\mathcal{P}$ into non-empty ordered words $\mathcal{Q}$ and $\mathcal{R}$. Some examples are
\begin{eqnarray}
    P = 132  \Rightarrow  (Q,R) &=& (1,32), (13,2)\nonumber\\
    \mathcal{P} = 123 \Rightarrow (\mathcal{Q},\mathcal{R}) &=& (12,3),(13,2),(23,1),\nonumber\\
    &&(1,23) ,(2,13), (3,12).\nonumber
\end{eqnarray}
The sums of the type  $\mathcal{Q}\cup \mathcal{R}= \mathcal{P}$  in the recursions for $K^{ak}$ and $\mathcal{K}^k$ arise from averaging over the permutation group, guaranteeing that the kernels are fully symmetric \cite{Mizera:2018jbh}.

The solution of  the recursion (\ref{Kcsrec})  for the color-stripped Yang-Mills kernel $K^k$ was found in \cite{Bardeen:1995gk,Cangemi:1996rx} and takes the beautiful and simple form
\begin{equation}\boxed{
         K^{ k}  (1 \ldots n) ={\delta^k_{1+ \ldots + n} \over S_{12} S_{23} \ldots S_{n-1,n}},\label{Kcssolgen} }
\end{equation}
where we have defined
\begin{equation}
     S_{12} := {s_{12} \over 2 \tilde F_{12}}= -{[1 2]\over  [q1][q2]}.\label{Sdef}
\end{equation}
To verify that this satisfies (\ref{Kcsrec}) one makes use of the identity
\begin{equation}
    2 \sum_{m=1}^{n-1} \tilde F_{1 + \ldots + m, m+1 + \ldots + n}S_{m,m+1} = s_{1\ldots n}.\label{Bardeenid}
\end{equation}
Furthermore, we have used that   $\Phi_{(1)}^k$ in (\ref{Kcsdef}-\ref{Kgravdef}) has support on the mass-shell $k^2=0$,  so that the momenta  $k_i$ can be treated as on-shell momenta and  the expression (\ref{XitoS}) for $\tilde F_{ij}$ can be used. One can also verify that the result (\ref{Kcssolgen}) satisfies the shuffle symmetry (\ref{Kcssymm}), making use of the properties
\begin{equation}
    S_{ij} + S_{ji}=0, \qquad S_{ij}+ S_{jk} + S_{ki}=0.
\end{equation}

Our goal will be to derive, in analogy with (\ref{Kcssolgen}),  similarly explicit solutions to the recursions (\ref{Kcdrec},\ref{Kgravrec}) for the color-dressed Yang-Mills kernel $K^{ak}$ and the gravity kernel $\mathcal{K}^k$. It is instructive to first work them out for $n = 2,3$. One finds that they can written suggestively as 
\begin{eqnarray}
       K^{ ak} (12 )&=& 
       = f_{a_1 a_2}^a  K^{ k} (1 2)\label{nis2solYM}\nonumber\\
        K^{ ak}  (1 23) 
        &=&   f_{a_1 a_2}^b  f_{b a_3}^a K^k (123) + f_{a_1 a_3}^b  f_{b a_2}^a K^k (132) \label{nis3solYM}\nonumber\\
          \mathcal{K}^{ k}  (12 )&=& 
          = F_{k_1 k_2}^k  \tilde  K  (1 2)\label{nis2solgrav}\\
       \mathcal{K}^{ k}  (1 23)
       &=&   F_{k_1 k_2}^l F_{l k_3}^k \tilde K  (123) + F_{k_1 k_3}^l  F_{l k_2}^k \tilde K  (132), \label{nis3sol}\nonumber\\
        \end{eqnarray}
where, as in (\ref{Fdef}), a tilde on a quantity means that we strip off an overall momentum-conserving delta function, e.g. 
$ K^k (1 \ldots n)=  \tilde K (1 \ldots n) \delta^k_{1+ \ldots +n}$ .  
In  (\ref{nis3sol}) we used the Jacobi identities for the  gauge and kinematic structure constants $f^a_{bc}$ resp.  $F^k_{lm}$. As it turns out, the  pattern of (\ref{nis3sol}) extends to all orders, and  the explicit solutions for the kernels  are
\begin{align}\boxed{\begin{aligned}
    K^{ ak}  (1 \ldots n)=&  \sum_{P \in S(2, \ldots ,n)}  c^a (\ell[1P])  K^k (1P) \\
           \mathcal{K}^{ k}  (1 \ldots n)=&  \sum_{P \in S(2, \ldots ,n)}  n^k (\ell[1P]) \tilde K (1P) .\label{Kgravgen}\end{aligned}}
\end{align}
Here,  the sum runs over all permutations of the letters $2, \ldots , n$ and  $c^a (\Gamma)$ and $n^k (\Gamma)$ are  color resp. kinematic group factors associated to a Lie monomial $\Gamma$, i.e. a fully bracketed expression of integer-valued letters. The definition $c^a (\Gamma)$ and $n^k (\Gamma)$ becomes clear upon considering some examples:

\begin{eqnarray}
        c^a( [[1,2],3]) &=&f_{a_1 a_2}^b  f_{ba_3}^a\nonumber\\
         n^k( [[1,2],3]) &=& F_{k_1 k_2}^l  F_{lk_3}^k \nonumber \\  c^a( [[1,2],[3,4]]) &= & f_{a1 a2}^b  f_{a_3a_4}^c f_{bc}^a \nonumber \\
        n^k( [[1,2],[3,4]]) &=&F_{k_1 k_2}^l  F_{k_3k_4}^m F_{lm}^k.    \nonumber
\end{eqnarray}
Finally, the notation $\ell [1\ldots n]$ stands for the left-bracketed Lie monomial
\begin{equation}
    \ell [1\ldots n] := [\ldots [[ 1,2],3]. \ldots n  ].\label{ls}
\end{equation}
More explicitly the group factors in (\ref{Kgravgen}) are given by
\begin{eqnarray}
    c^a (\ell[1 \ldots n]) &=& f_{12}^{b_1} f_{b_1 3}^{b_2} \ldots f_{b_{n-2} n}^a \nonumber \\
     n^k (\ell[1 \ldots n]) &=& F_{12}^{l_1} F_{l_1 3}^{l_2} \ldots F_{l_{n-2} n}^k\nonumber \\
       &=& \delta^k_{1+\ldots +n}\sum_{\{a_i\}, a_i\leq i } { \tilde F_{12 }\tilde F_{a_2 3 } \ldots \tilde F_{a_{n-1} n }}\nonumber 
\end{eqnarray}
Note in particular that the kinematic factors $ n^k (\ell[1 \ldots n])$ are local, i.e. without poles in the momenta, since the $\tilde F_{ij}$ are local, see  (\ref{XitoS}). From this and  the  $K^k$ momentum dependence   one sees  that  the momentum integrals in (\ref{Kcsdef}-\ref{Kgravdef}) do not suffer from soft divergences.    In Appendix \ref{Appproof} we prove that the expressions (\ref{Kgravgen}) solve the recursions (\ref{Kcdrec},\ref{Kgravrec}). The proof relies on elementary properties of Lie polynomials which are conveniently reviewed in \cite{Frost:2020eoa}.  The result (\ref{Kgravgen})  shows that the general perturbative solution to the selfdual Yang-Mills and gravity equations can be written in a BCJ-like form and are  related by color-kinematics duality.

\section{Double-copy form}

In  this section we rewrite our formula for the selfdual gravity solution  in (\ref{Kgravgen})  in a manifestly double-copy form. To achieve this, we first express     (\ref{Kgravgen})  in terms of combinatorial objects which were introduced in  \cite{Mafra:2015vca, Mafra:2020qst,Frost:2020eoa} (see also \cite{Escudero:2022zdz,Correa:2024mub}) to solve recursion relations for  Berends-Giele currents   using the theory of Lie polynomials.   
Having obtained the double-copy form of the general classical solution, it is straightforward to derive from it   double copy relations for tree  level amplitudes, as  will be discussed in the next section.

We start by introducing the following hierarchy of combinatorial objects:
\begin{eqnarray}
    b &=& \sum_\Gamma {\Gamma\otimes\Gamma\over s_\Gamma} \label{bdef}\\
    b(P)  &=& \sum_\Gamma {(P,\Gamma) \Gamma\over s_\Gamma}\label{bPdef}\\
      b(P|Q)  &=& \sum_\Gamma {(P,\Gamma) (Q,\Gamma)\over s_\Gamma} \label{bPQdef}
\end{eqnarray}
Here, the sum runs over all Lie monomials $\Gamma$, and the inner product of words is defined as $(P,Q)= \delta_{P,Q}$ and extended to combinations of words by linearity. 

Let us first discuss how these objects encode the combinatorial information contained in  the integration kernels introduced in the previous section. We start  with the object $b(P)$ in (\ref{bPdef}), which can be seen to generate the binary tree graphs arising in the diagrammatic solution of the quadratic recursion (\ref{Kcsrec}).
In particular it can be shown  \cite{Frost:2020eoa} that $b(q)$  satisfies the recursion relation
\begin{equation}
    b(P) = \sum_{QR =P} [b(Q), b(R)], \qquad b(i) = 1.
\end{equation}
Comparing to (\ref{Kcsrec}) shows  that $b (P)$ is related to the color-stripped kernel as
\begin{equation}
    K^k (P) =2^{|P|-1} n^k (b(P)).\label{KcditobP}
\end{equation}
Now let us discuss the meaning of $b$ in (\ref{bdef}).  A basis of   Lie monomials is furnished by the  left-bracketed monomials (\ref{ls}), in terms of which any Lie polynomial of length $n$ can be expanded as
\begin{equation}
    \Gamma = \sum_{P \in S(2, \ldots,n)} (1P,\Gamma) \ell[1P].
\end{equation}
We see from this that our expressions (\ref{Kgravgen}) can be written as
\begin{eqnarray}
      K^{ak}  (1 \ldots n) &=& 2^{n-1} (c^a  \otimes n^k )  ( b), \\
    \mathcal{K}^k  (1 \ldots n) &=& 2^{n-1} (  \tilde  n   \otimes n^k )( b ),
\end{eqnarray}
where we have defined e.g. $(c^a  \otimes n^k) (\Gamma_1 \otimes \Gamma_2) = c^a (\Gamma_1) n^k (\Gamma_2 )  $. This shows that $b$  is the basic combinatorial object underlying the color-dressed and gravity kernels. 
The object $ b(P|Q) $ in (\ref{bPQdef}) can be shown to govern the  color-stripped kernel of a biadjoint scalar theory, which we will not discuss here.  

To write the gravity kernel in a double-copied form, we first perform  a basis expansion of the Lie polynomial  $b(P)$ in (\ref{KcditobP}):
\begin{equation}
     K^k  (1P) = 2^{ |P|}\sum_{Q\in S(2, \ldots , n)} b(1P|1Q) n^k  ( \ell[1Q] ) 
\end{equation}
As shown in \cite{Frost:2020eoa}, this relation can be inverted to obtain the kinematic numerator $ n^p  ( \ell[1Q] ) $ by means  the KLT matrix $S(P|Q)_1$ \cite{Bern:1998sv,Bjerrum-Bohr:2010diw}:
\begin{equation}
     n^k  ( \ell[1P] ) = 2^{ -|P|} \sum_{Q\in S(2, \ldots , n)}  S(P|Q)_1  K^k  (1Q).
\end{equation}
Substituting in (\ref{Kgravgen}) we obtain a double-copy form for the gravity kernel, expressed in terms two copies of the color-stripped Yang-Mills kernel and the KLT matrix:
\begin{equation}\boxed{
   \tilde{ \mathcal{K}}  (1 \ldots n) = 2^{1-n}\sum_{P,Q \in S (2\ldots n)}\tilde K  (1P)  S(P|Q)_1 \tilde K  (1P) .\label{KDC}
}\end{equation}
To make the expression fully explicit, we recall from (\ref{Kcssolgen}, \ref{Sdef}) that
\begin{equation}
\tilde  K  (1\ldots n)     = (-1)^{ n-1} {[q1][q2]^2 \ldots [q (n-1)]^2[qn] \over 
 	[12][23]\ldots [(n-1)n]}
\end{equation}
and that the KLT matrix is given by \cite{Bjerrum-Bohr:2010diw}
\begin{equation}
    S[1_1 \ldots i_k|j_1 \ldots j_k]_1 = \prod_{t=1}^k(s_{i_t 1} + \sum_{q>k} \theta (i_t, i_q) s_{i_t i_q}),
\end{equation}
where $\theta (i_t, i_q)$ is 0 if $i_t$ sequentially comes before $i_q$ in $\{j_1, \ldots , j_k\}$, and 1 otherwise.  

For example, for $n=3$ the formula (\ref{KDC}) yields
\begin{widetext}
\begin{eqnarray}
\tilde{ \mathcal{K}}  (1 23) &=&    {1\over 4} \left(  \tilde K{(123)} \tilde K{(123)} ( s_{12} s_{13} +  s_{12} s_{23}) + \tilde K{(123)} \tilde K{(132)} s_{12} s_{13} + \tilde K{(132)} \tilde K{(123)} s_{12} s_{13} \right.\nonumber \\ && \left.+ \tilde K{(132)} \tilde K{(132)} ( s_{13} s_{12} +  s_{13} s_{23})\right)\nonumber\\
&=&  {1\over 4} \left((\tilde K (123))^2 s_{12} s_{23} +  (\tilde K (312))^2 s_{31} s_{12}+  (\tilde K (231))^2 s_{23} s_{31}\right), 
\end{eqnarray}
\end{widetext}
where in the last identity we used the shuffle symmetry $\tilde K (12 \shuffle 3 )=0$. It's straightfroward to show that this last expression is equivalent to (\ref{nis3sol}).
\section{Amplitude double copy relations }
Having derived that perturbative classical solutions exhibit a double-copy structure,  we now directly  link this  to  double-copy relations for amplitudes. The cleanest link between equations of motion and tree-level amplitudes is through the perturbiner expansion \cite{Rosly:1996cp,Rosly:1996vr,Selivanov:1997aq,Rosly:1997ap} (see \cite{Mizera:2018jbh, Gomez:2021shh} for a modern perspective), which yields  the Berends-Giele (BG) currents of any field  theory from a perturbative solution to its equations of motion. The BG currents are  tree-level form factors, i.e.  amplitudes in which one particle is off-shell while all the others are on-shell,  from which on-shell amplitudes are easily derived.  We now show that our above results lead directly to known double copy relations both for BG currents and on-shell amplitudes.

The perturbiner expansions  for the selfdual Yang-Mills and gravity prepotentials  take the form
\begin{eqnarray}
    \Phi^k &=& \sum_P g^{|P|-1} J_P T^P \delta^k_{p_P}\nonumber\\
    &=& \sum_i J_i T^{a_i} \delta^k_{p_i} + g \sum_{i,j}J_{ij} T^{a_i}T^{a_j} \delta^k_{p_{ij}}+\ldots
    \end{eqnarray}
    \begin{eqnarray}
     \Psi^k &=& \sum_\mathcal{P} \left( {\kappa \over 2} \right)^{|\mathcal{P}|-1}\mathcal{J}_\mathcal{P}  \delta^k_{p_\mathcal{P}}\nonumber\\
     &=& \sum_i \mathcal{J}_i   \delta^k_{p_i} + {\kappa \over 2}\sum_{i < j}\mathcal{J}_{ij}  \delta^k_{p_{ij}}+\ldots\label{perturbpot}
\end{eqnarray}
Here, the momenta $p_i$ are on-shell, $p_i^2$=0, and the coefficients $J_i, J_{ij}, \mathcal{J}_i, \mathcal{J}_{ij},\ldots$ are assumed to be nilpotent,  $ J_i^2= J_{ij}^2 = \mathcal{J}_i^2= \mathcal{J}_{ij}^2 =\ldots =0$,  in order to omit terms which do not contribute to the BG currents. The expansions (\ref{perturbpot}) give rise to perturbiner expansions of the gauge potential and the metric  through the relations (\ref{AitoF},\ref{hitoPsi}):
\begin{equation}
    A^k_{\alpha\dot \alpha} = \sum_P g^{|P|}A^P_{\alpha\dot \alpha} T^P \delta^k_{p_P}, \ \ \ h^k_{\alpha\dot\alpha\beta\dot\beta } = \sum_\mathcal{P} {\kappa^{|\mathcal{P}|}\over 2 ^{|\mathcal{P}|-1}}h^\mathcal{P}_{\alpha\dot\alpha\beta\dot\beta }\delta^k_{p_\mathcal{P}},
\end{equation}
with 
\begin{equation}
    A^P_{\alpha\dot \alpha}=  v_{\alpha \dot \alpha}^P  J_P,\qquad  h^\mathcal{P}_{\alpha\dot\alpha\beta\dot\beta } = v_{\alpha \dot \alpha}^\mathcal{P} v_{\beta \dot \beta}^\mathcal{P}\mathcal{J}_\mathcal{P}
\end{equation}
and where
\begin{equation}
    v_{\alpha \dot \alpha}^P  = i\mathfrak{q}_\alpha \sum_{i=1}^{|P|}  [q P_i]\tilde{\mathfrak{p}}_{\dot \alpha}^{P_i} .
\end{equation}

Evaluating  $\Phi_{(n)}^k, \Psi_{(n)}^k$ in (\ref{Kcsdef},\ref{Kgravdef}) with the first order parts given by 
\begin{equation}
        \Phi_{(1)}^k=  \sum_i J_i T^{a_i} \delta^k_{p_i} , \qquad   \Psi_{(1)}^k =\sum_i \mathcal{J}_i   \delta^k_{p_i} 
\end{equation}
we readily find the currents in terms of the integration kernels
\begin{eqnarray}
       J_{i_1\ldots i_n} &=& J_{i_1} \ldots J_{i_n} \tilde K_{k_{i_1} \ldots k_{i_n}}\\ \mathcal{J}_{i_1\ldots i_n} &=& \mathcal{J}_{i_1} \ldots \mathcal{J}_{i_n} \tilde{\mathcal{K}}_{k_{i_1} \ldots k_{i_n}}.\label{currents}
\end{eqnarray}
We should remark that, while the kernels  $ \tilde K_{k_{1} \ldots k_{n}} $ and $\tilde{\mathcal{K}}_{k_{1} \ldots k_{n}}$ uniquely determine the  currents, the converse is not true: due to the nilpotency of the $J_i$ and $\mathcal{J}_i$  we could add arbitrary  `contact terms'
with delta-function support on  the locus where some  of the momenta being equal, without changing the currents. In the selfdual case at hand, we  have demonstrated that such contact terms are in fact absent and the kernels are simple analytic continuations of the currents; it would be interesting to understand when this is the case in general.

The  canonically normalized BG currents  are obtained from (\ref{currents}) by  treating $J_i$ and  $\mathcal{J}_i$  now as c-numbers and substituting the appropriate values  (\ref{Jvals}) , i.e.
\begin{equation}
    J_i = - {i \over [qi]^2}, \qquad \mathcal{J}_i = - {1 \over [qi]^4}.
\end{equation}
Upon making this substitution the BG currents compute the off-shell tree-level amplitudes
\begin{eqnarray}
  \langle 1^+ \ldots n^+ A_{\alpha\dot \alpha} (k) \rangle_{\rm tree}  &=&   A_{\alpha\dot \alpha} ^{1 \ldots n} \delta^k_{1 \ldots n}\\
    \langle 1^{++} \ldots n^{++} h_{\alpha\dot \alpha\beta\dot \beta} (k) \rangle_{\rm tree}  &=&   h_{\alpha\dot \alpha\beta\dot \beta}^{1 \ldots n} \delta^k_{1 \ldots n}\label{BGampl}
\end{eqnarray}
From (\ref{KDC}) we obtain a double-copy relation for BG currents
\begin{equation}
\boxed{     h_{\alpha \dot \alpha \beta \dot \beta}^{1\ldots n} =2^{1-n} \sum_{P,Q \in S(2,\ldots ,n)} A_{\alpha\dot \alpha}^{1 P}  S [P|Q]_1  A_{\beta\dot \beta}^{1 Q}.\label{gravBG}}
\end{equation}
This relation was conjectured in \cite{Cho:2021nim} and recently proven  in  \cite{Correa:2024mub} using results from \cite{Escudero:2022zdz}.  Once again we can obtain a more explicit formula by substituting the Yang-Mills currents 

\begin{equation}
     A_{\alpha\dot \alpha}^{1 \ldots n} =  {i^{n+1} \mathfrak{q}_\alpha \sum_{i=1}^{n}  [q i] \tilde{\mathfrak{p}}_{\dot \alpha}^{i}
   \over [12][23] \ldots [(n-1)n] [q1] [qn]}.\label{YMBG}
\end{equation}
We note that this  expression agrees with the one found in \cite{Rosly:1996vr} using somewhat different methods..

Let us now discuss the double copy relations for on-shell amplitudes which follow from the above.  Equation  (\ref{BGampl}) implies that from the selfdual BG currents we can compute the on-shell  amplitudes 
\begin{widetext}
\begin{eqnarray}
  \langle 1^+ \ldots n^+ (n+1)^\pm \rangle_{\rm tree}  &=&   \lim_{s_{1 \ldots n} \to 0} s_{1 \ldots n}  A_{\alpha\dot \alpha} ^{1 \ldots n} e_\pm^{\alpha \dot \alpha} (p_{n+1})\delta^{n+1}_{1 \ldots n}\nonumber\\
  \langle 1^{++} \ldots n^{++} (n+1)^{\pm \pm} \rangle_{\rm tree}  &=&  \lim_{s_{1 \ldots n} \to 0} s_{1 \ldots n}  h_{\alpha\dot \alpha\beta\dot \beta}^{1 \ldots n} e_{\pm\pm}^{\alpha \dot \alpha\beta\dot \beta} (p_{n+1})\delta^{n+1}_{1 \ldots n}\nonumber
\end{eqnarray}\end{widetext}
These can be evaluated using (\ref{pols},\ref{gravBG},\ref{YMBG}).   For $n>2$, both the Yang-Mills and gravity amplitudes vanish due to the absence of a pole in $s_{1 \ldots n} $ in the combinations on the RHS.  For $n=2$ however we find the standard nonzero results
\begin{eqnarray}
    \langle 1^+ 2^+ 3^- \rangle &=& i{\langle 12 \rangle [q3]^2 \over [q1][q2]}= - i {\langle 12 \rangle^3  \over \langle 13 \rangle \langle 23 \rangle}\\
    \langle 1^{++} 2^{++} 3^{--} \rangle &=& \langle 1^+ 2^+ 3^- \rangle^2\label{DConshell}
\end{eqnarray}
In the first line we have used momentum-conservation relations such as $[q3]\langle 23\rangle= - [q1]\langle 21\rangle$, while the second line readily follows from (\ref{gravBG}) and the fact that $S(2|2)_1 = s_{12}$. 

To summarize, though the selfdual classical double copy (\ref{Kgravgen}) leads to infinitely many nontrivial double copy relations between BG currents (\ref{gravBG}) ,  the three-point double copy relation (\ref{DConshell}) is the  only nonvanishing on-shell  amplitude relation  which is directly encoded  in it. Nevertheless, the appearance of the KLT matrix, which governs double copy relations of general amplitudes, in (\ref{Kgravgen})  suggests that selfdual solutions `know' about more general on-shell amplitudes than (\ref{DConshell}). Indeed, more general amplitudes can be obtained by perturbing around (\ref{Kgravgen})  away from selfduality \cite{Miller:2024oza}.

\section{Discussion and outlook}
In this letter we have derived color-kinematics duality  (\ref{Kgravgen}) and double-copy relations (\ref{KDC}) in the  perturbative structure of selfdual solutions.  Let us first comment on  the connection with the more standard viewpoint of the classical double copy as an explicit map between gauge and gravity solutions. A canonical map between solutions  exists only in the case where the gauge theory solution lives in a  $u(1)$ subalgebra (and therefore linearizes), $\Phi =  \Phi_{(1)} c^a T_a$. Then the unique double copied solution is obtained from the linearized staring point  $\Psi_{(1)}= \Phi_{(1)}$. We note that this gravity solution doesn't linearize in general. Most of the literature focuses on special examples where the gravity solution itself also linearizes, such as the pp-wave  \cite{Monteiro:2014cda} 
$\Psi_{(1)}= \Phi_{(1)}\sim e^{i p x}$ and the Eguchi-Hanson   \cite{Berman:2018hwd}  $\Psi_{(1)}= \Phi_{(1)}\sim{1 \over x^2}$ solutions.  For genuinely non-abelian Yang-Mills solutions, there is in general no unique  double copied gravity solution, since we can pick the component $\Phi_{(1)}^a$ along any color direction $T_a$ to be the  gravity perturbation $\Psi_{(1)}$. 

In spite of this, one may wonder  whether a truly nonabelian double copy might be defined for certain subclasses  of Yang-Mills solutions. For this it would be necessary to  embed their (kinematic  $\oplus$   gauge) structure in the gravitational kinematic algebra $sdiff_2$.  For example, if the gauge group is $SU(2)$ the   generators  can be embedded in the kinematic algebra as  (complex) combinations of $w^{(2)}_1,w^{(2)}_0,w^{(2)}_{-1}$ given in (\ref{Winftgenrs}).  A Yang-Mills solution with first order components $\Phi_{(1)}^m$ can then be tentatively be double copied to a gravity solution with first order  part $\Psi_{(1)} =  \Phi_{(1)}^m w^{(2)}_m$. For this to make sense  we have to further impose $\partial_{\alpha \dot \alpha} \Phi_{(1)}^m  \partial^{\alpha \dot \alpha}  w^{(2)}_m =0$. At second order, the structure of the YM solution (\ref{SDYM}) is reproduced if we impose $\{ \partial_{\dot \alpha }\Psi_{(1)}, \partial^{\dot \alpha }\Psi_{(1)}\} = 2 \{  \Phi_{(1)}^m, \Phi_{(1)}^n\}  \{w^{(2)}_m,w^{(2)}_n\}$, and so on. It would be interesting to explore if this procedure can  lead to nontrivial solutions.

Let us comment on some further open issues. Though the selfdual theories we have considered are integrable, this structure was not used directly in our approach. It would therefore be interesting to make contact with integrability techniques such as Bethe ans\"atze as well as with the twistor space construction  for selfdual gravity solutions \cite{Penrose:1976js} which underly the form of the selfdual perturbiner solution presented in \cite{Rosly:1997ap}. Finally, it would be of obvious interest  to extend the present work beyond the selfdual sector.

\begin{acknowledgments}
    It's a pleasure  to thank  Vojtech Pravda for a useful discussion and Renann Lipinski Jusinskas and Cristhiam Lopez-Arcos for patient explanations of the perturbiner method and for   comments on the manuscript. This work was supported by the Grant Agency of the Czech Republic under the grant
EXPRO 20-25775X.
\end{acknowledgments} 

\begin{appendix}
    \section{Proof of the solution to the recursion relations}\label{Appproof}
In this Appendix, we will prove that the expression  (\ref{Kgravgen}) for the selfdual gravity kernel,
\begin{equation}
     \mathcal{K}^{ k}  (1 \ldots n)= \sum_{P \in S(2, \ldots ,n)}  n^k (\ell[1P]) \tilde K (1P) ,\label{Kgravgenapp}
\end{equation}
solves the recursion (\ref{Kgravrec}),
\begin{equation}
    \mathcal{K}^{ k} (1 \ldots n) = {1 \over   s_{(1 \ldots n)}} \sum_{\mathcal{Q}\cup \mathcal{R}=1\ldots n } {\tilde F_{lm} F^k_{lm}}    \mathcal{K}^{ l} (\mathcal{Q}) \mathcal{K}^{ m}    (\mathcal{R}) \label{Kgravrecapp}
    \end{equation}
The proof that the  color-dressed Yang-Mills kernel  in  (\ref{Kgravgen}) solves  (\ref{Kcdrec}) is  
completely analogous under replacing kinematic structure constants with gauge ones.  We will work by induction in $n$.  For $n=2$,   the solution of the recursion (\ref{nis2solgrav}) agrees with (\ref{Kgravgenapp}). Assuming that $\mathcal{K}^{ k}  (1 \ldots m)$  given by  (\ref{Kgravgenapp}) is correct for all $m <n$, we write the RHS of (\ref{Kgravrecapp}
) as
\begin{equation}
    {1 \over   s_{(1 \ldots n)}} \sum_{1\ldots n =\mathcal{Q}\cup \mathcal{R}} {\tilde F_{lm} F^k_{lm}}    \mathcal{K}^{ l} (\mathcal{Q}) \mathcal{K}^{ m}    (\mathcal{R}) = n^k (\Lambda),
\end{equation}
where $\Lambda$ is the following Lie polynomial with momentum-dependent coefficients
\begin{widetext}\begin{equation}
    \Lambda =  {1 \over   s_{(1 \ldots n)}} \sum_{\mathcal{Q}\cup \mathcal{R}=1\ldots n }  \sum_{\begin{array}{l} Q \in S( \mathcal{Q}_2,  \ldots, \mathcal{Q}_{|\mathcal{Q}|} ) \\ 
    R \in S( \mathcal{R}_2,  \ldots, \mathcal{R}_{|\mathcal{R}|} )  \end{array}}[\ell[\mathcal{Q}_1 Q ], \ell[\mathcal{R}_1 R ]]   \tilde F_{lm} K^l ( \mathcal{Q}_1 Q ) K^m ( \mathcal{R}_1 R ).
\end{equation}\end{widetext}
Next, we expand $\Lambda$ in the basis of left-bracketed monomials as
\begin{equation}
    \Lambda = \sum_{P \in S(2, \ldots , n)} (1P, \Lambda) \ell[1,P],
\end{equation}
where the inner product of words is defined as $(P,Q) \equiv \delta_{P,Q}$ and is extended to  Lie polynomials by linearity. To prove (\ref{Kgravgenapp}) it remains to show that $(1P,\Lambda) = \tilde K (1P)$, which 
proceeds in the following steps:
\begin{widetext}
\begin{eqnarray}
    (1P,\Lambda) &=&{1\over   s_{(1 \ldots n)}}  \sum_{XY=1P }\sum_{\mathcal{Q}\cup \mathcal{R}=1\ldots n }  \sum_{\begin{array}{l} Q \in S( \mathcal{Q}_2,  \ldots, \mathcal{Q}_{|\mathcal{Q}|} ) \nonumber \\
    R \in S( \mathcal{R}_2,  \ldots, \mathcal{R}_{|\mathcal{R}|} )  \end{array}} \left( (X,\ell[\mathcal{Q}_1 Q ])( Y,\ell[\mathcal{R}_1 R ]) -  (Y,\ell[\mathcal{Q}_1 Q ])( X,\ell[\mathcal{R}_1 R ])\right)\nonumber\\
&&   \tilde F_{lm} K^l( \mathcal{Q}_1 Q ) K^m ( \mathcal{R}_1 R )   \nonumber \\
     &=&  {2\over   s_{(1 \ldots n)}}\sum_{ XY=1P } \tilde F_{lm} K^l (X) K^m \left(\sum_{  R \in S( Y \backslash Y_{min} )} \left(Y, \ell[Y_{min} R]\right) Y_{min} R\right)\nonumber \\
     &=&  {2\over   s_{(1 \ldots n)}}  \sum_{ 1P=XY }\tilde F_{lm} K^l (X) K^m (Y)\nonumber\\
     &=& \tilde K (1P)\qquad \qquad \blacksquare \nonumber
\end{eqnarray}
\end{widetext}
In the first equality, we have used the identity (see eq.  (A.10) in \cite{Frost:2020eoa})
\begin{equation}
   (Z,[\Gamma_1, \Gamma_2] )  = \sum_{XY = Z}\left( (X,\Gamma_1) (Y,\Gamma_2) - (X,\Gamma_2) (Y,\Gamma_1) \right).
\end{equation}
In the second equality, we have used that the first letter of $X$ is 1, so that $ (X,\ell[\mathcal{Q}_1 Q ])$ is nonvanishing only when $\mathcal{Q}_1 =1$ (since $\mathcal{Q}_1 < Q_i , \,\forall i$). Furthermore, from the identity $(1 Z,\ell[1 T ]) = \delta_{Z,T}$ we see that the $(X,\ell[\mathcal{Q}_1 Q ])( Y,\ell[\mathcal{R}_1 R ]) $  terms in the sum are nonvanishing only when  $\mathcal{Q}_1 Q = X$ and when $\mathcal{R}$ is  obtained from $Y$ by  reordering its letters  in increasing order.    We denoted by $Y_{min}$ the smallest letter in $Y$. 
The $(Y,\ell[\mathcal{Q}_1 Q ])( X,\ell[\mathcal{R}_1 R ]) $ terms give the same contribution leading to the factor of 2. In the third equality we have used that (see eq. (2.10) in \cite{Frost:2020eoa})
\begin{equation}
    \sum_{T \in S(2, \ldots, n) }( Y, \ell[1T]) 1T = Y + {\rm complete\  shuffles},
\end{equation}
(where the omitted terms are combinations of complete shuffles $A \shuffle B$), as well as the shuffle symmetry of $K^l$  (\ref{Kcssymm}). Finally, in the last equality we used the recursion relation for $\tilde K$ which follows from (\ref{Kcsrec}). 

\end{appendix}

\bibliography{refsCDC}

\begin{thebibliography}{39}%
\makeatletter
\providecommand \@ifxundefined [1]{%
 \@ifx{#1\undefined}
}%
\providecommand \@ifnum [1]{%
 \ifnum #1\expandafter \@firstoftwo
 \else \expandafter \@secondoftwo
 \fi
}%
\providecommand \@ifx [1]{%
 \ifx #1\expandafter \@firstoftwo
 \else \expandafter \@secondoftwo
 \fi
}%
\providecommand \natexlab [1]{#1}%
\providecommand \enquote  [1]{``#1''}%
\providecommand \bibnamefont  [1]{#1}%
\providecommand \bibfnamefont [1]{#1}%
\providecommand \citenamefont [1]{#1}%
\providecommand \href@noop [0]{\@secondoftwo}%
\providecommand \href [0]{\begingroup \@sanitize@url \@href}%
\providecommand \@href[1]{\@@startlink{#1}\@@href}%
\providecommand \@@href[1]{\endgroup#1\@@endlink}%
\providecommand \@sanitize@url [0]{\catcode `\\12\catcode `\$12\catcode `\&12\catcode `\#12\catcode `\^12\catcode `\_12\catcode `\%12\relax}%
\providecommand \@@startlink[1]{}%
\providecommand \@@endlink[0]{}%
\providecommand \url  [0]{\begingroup\@sanitize@url \@url }%
\providecommand \@url [1]{\endgroup\@href {#1}{\urlprefix }}%
\providecommand \urlprefix  [0]{URL }%
\providecommand \Eprint [0]{\href }%
\providecommand \doibase [0]{https://doi.org/}%
\providecommand \selectlanguage [0]{\@gobble}%
\providecommand \bibinfo  [0]{\@secondoftwo}%
\providecommand \bibfield  [0]{\@secondoftwo}%
\providecommand \translation [1]{[#1]}%
\providecommand \BibitemOpen [0]{}%
\providecommand \bibitemStop [0]{}%
\providecommand \bibitemNoStop [0]{.\EOS\space}%
\providecommand \EOS [0]{\spacefactor3000\relax}%
\providecommand \BibitemShut  [1]{\csname bibitem#1\endcsname}%
\let\auto@bib@innerbib\@empty
\bibitem [{\citenamefont {Bern}\ \emph {et~al.}(2024)\citenamefont {Bern}, \citenamefont {Carrasco}, \citenamefont {Chiodaroli}, \citenamefont {Johansson},\ and\ \citenamefont {Roiban}}]{Bern:2019prr}%
  \BibitemOpen
  \bibfield  {author} {\bibinfo {author} {\bibfnamefont {Z.}~\bibnamefont {Bern}}, \bibinfo {author} {\bibfnamefont {J.~J.}\ \bibnamefont {Carrasco}}, \bibinfo {author} {\bibfnamefont {M.}~\bibnamefont {Chiodaroli}}, \bibinfo {author} {\bibfnamefont {H.}~\bibnamefont {Johansson}},\ and\ \bibinfo {author} {\bibfnamefont {R.}~\bibnamefont {Roiban}},\ }\bibfield  {title} {\bibinfo {title} {{The duality between color and kinematics and its applications}},\ }\href {https://doi.org/10.1088/1751-8121/ad5fd0} {\bibfield  {journal} {\bibinfo  {journal} {J. Phys. A}\ }\textbf {\bibinfo {volume} {57}},\ \bibinfo {pages} {333002} (\bibinfo {year} {2024})},\ \Eprint {https://arxiv.org/abs/1909.01358} {arXiv:1909.01358 [hep-th]} \BibitemShut {NoStop}%
\bibitem [{\citenamefont {Monteiro}\ \emph {et~al.}(2014)\citenamefont {Monteiro}, \citenamefont {O'Connell},\ and\ \citenamefont {White}}]{Monteiro:2014cda}%
  \BibitemOpen
  \bibfield  {author} {\bibinfo {author} {\bibfnamefont {R.}~\bibnamefont {Monteiro}}, \bibinfo {author} {\bibfnamefont {D.}~\bibnamefont {O'Connell}},\ and\ \bibinfo {author} {\bibfnamefont {C.~D.}\ \bibnamefont {White}},\ }\bibfield  {title} {\bibinfo {title} {{Black holes and the double copy}},\ }\href {https://doi.org/10.1007/JHEP12(2014)056} {\bibfield  {journal} {\bibinfo  {journal} {JHEP}\ }\textbf {\bibinfo {volume} {12}},\ \bibinfo {pages} {056}},\ \Eprint {https://arxiv.org/abs/1410.0239} {arXiv:1410.0239 [hep-th]} \BibitemShut {NoStop}%
\bibitem [{\citenamefont {White}(2024)}]{White:2024pve}%
  \BibitemOpen
  \bibfield  {author} {\bibinfo {author} {\bibfnamefont {C.~D.}\ \bibnamefont {White}},\ }\href {https://doi.org/10.1142/q0457} {\emph {\bibinfo {title} {{The Classical Double Copy}}}}\ (\bibinfo  {publisher} {World Scientific},\ \bibinfo {year} {2024})\BibitemShut {NoStop}%
\bibitem [{\citenamefont {Luna}\ \emph {et~al.}(2016)\citenamefont {Luna}, \citenamefont {Monteiro}, \citenamefont {Nicholson}, \citenamefont {O'Connell},\ and\ \citenamefont {White}}]{Luna:2016due}%
  \BibitemOpen
  \bibfield  {author} {\bibinfo {author} {\bibfnamefont {A.}~\bibnamefont {Luna}}, \bibinfo {author} {\bibfnamefont {R.}~\bibnamefont {Monteiro}}, \bibinfo {author} {\bibfnamefont {I.}~\bibnamefont {Nicholson}}, \bibinfo {author} {\bibfnamefont {D.}~\bibnamefont {O'Connell}},\ and\ \bibinfo {author} {\bibfnamefont {C.~D.}\ \bibnamefont {White}},\ }\bibfield  {title} {\bibinfo {title} {{The double copy: Bremsstrahlung and accelerating black holes}},\ }\href {https://doi.org/10.1007/JHEP06(2016)023} {\bibfield  {journal} {\bibinfo  {journal} {JHEP}\ }\textbf {\bibinfo {volume} {06}},\ \bibinfo {pages} {023}},\ \Eprint {https://arxiv.org/abs/1603.05737} {arXiv:1603.05737 [hep-th]} \BibitemShut {NoStop}%
\bibitem [{\citenamefont {Goldberger}\ and\ \citenamefont {Ridgway}(2017)}]{Goldberger:2016iau}%
  \BibitemOpen
  \bibfield  {author} {\bibinfo {author} {\bibfnamefont {W.~D.}\ \bibnamefont {Goldberger}}\ and\ \bibinfo {author} {\bibfnamefont {A.~K.}\ \bibnamefont {Ridgway}},\ }\bibfield  {title} {\bibinfo {title} {{Radiation and the classical double copy for color charges}},\ }\href {https://doi.org/10.1103/PhysRevD.95.125010} {\bibfield  {journal} {\bibinfo  {journal} {Phys. Rev. D}\ }\textbf {\bibinfo {volume} {95}},\ \bibinfo {pages} {125010} (\bibinfo {year} {2017})},\ \Eprint {https://arxiv.org/abs/1611.03493} {arXiv:1611.03493 [hep-th]} \BibitemShut {NoStop}%
\bibitem [{\citenamefont {Bautista}\ and\ \citenamefont {Guevara}(2019)}]{Bautista:2019tdr}%
  \BibitemOpen
  \bibfield  {author} {\bibinfo {author} {\bibfnamefont {Y.~F.}\ \bibnamefont {Bautista}}\ and\ \bibinfo {author} {\bibfnamefont {A.}~\bibnamefont {Guevara}},\ }\bibfield  {title} {\bibinfo {title} {{From Scattering Amplitudes to Classical Physics: Universality, Double Copy and Soft Theorems}},\ }\href@noop {} {\  (\bibinfo {year} {2019})},\ \Eprint {https://arxiv.org/abs/1903.12419} {arXiv:1903.12419 [hep-th]} \BibitemShut {NoStop}%
\bibitem [{\citenamefont {Shen}(2018)}]{Shen:2018ebu}%
  \BibitemOpen
  \bibfield  {author} {\bibinfo {author} {\bibfnamefont {C.-H.}\ \bibnamefont {Shen}},\ }\bibfield  {title} {\bibinfo {title} {{Gravitational Radiation from Color-Kinematics Duality}},\ }\href {https://doi.org/10.1007/JHEP11(2018)162} {\bibfield  {journal} {\bibinfo  {journal} {JHEP}\ }\textbf {\bibinfo {volume} {11}},\ \bibinfo {pages} {162}},\ \Eprint {https://arxiv.org/abs/1806.07388} {arXiv:1806.07388 [hep-th]} \BibitemShut {NoStop}%
\bibitem [{\citenamefont {P.~V.}\ and\ \citenamefont {Manu}(2020)}]{PV:2019uuv}%
  \BibitemOpen
  \bibfield  {author} {\bibinfo {author} {\bibfnamefont {A.}~\bibnamefont {P.~V.}}\ and\ \bibinfo {author} {\bibfnamefont {A.}~\bibnamefont {Manu}},\ }\bibfield  {title} {\bibinfo {title} {{Classical double copy from Color Kinematics duality: A proof in the soft limit}},\ }\href {https://doi.org/10.1103/PhysRevD.101.046014} {\bibfield  {journal} {\bibinfo  {journal} {Phys. Rev. D}\ }\textbf {\bibinfo {volume} {101}},\ \bibinfo {pages} {046014} (\bibinfo {year} {2020})},\ \Eprint {https://arxiv.org/abs/1907.10021} {arXiv:1907.10021 [hep-th]} \BibitemShut {NoStop}%
\bibitem [{\citenamefont {Lee}(2022)}]{Lee:2022aiu}%
  \BibitemOpen
  \bibfield  {author} {\bibinfo {author} {\bibfnamefont {K.}~\bibnamefont {Lee}},\ }\bibfield  {title} {\bibinfo {title} {{Quantum off-shell recursion relation}},\ }\href {https://doi.org/10.1007/JHEP05(2022)051} {\bibfield  {journal} {\bibinfo  {journal} {JHEP}\ }\textbf {\bibinfo {volume} {05}},\ \bibinfo {pages} {051}},\ \Eprint {https://arxiv.org/abs/2202.08133} {arXiv:2202.08133 [hep-th]} \BibitemShut {NoStop}%
\bibitem [{\citenamefont {Gomez}\ \emph {et~al.}(2023)\citenamefont {Gomez}, \citenamefont {Lipinski~Jusinskas}, \citenamefont {Lopez-Arcos},\ and\ \citenamefont {Quintero~Velez}}]{Gomez:2022dzk}%
  \BibitemOpen
  \bibfield  {author} {\bibinfo {author} {\bibfnamefont {H.}~\bibnamefont {Gomez}}, \bibinfo {author} {\bibfnamefont {R.}~\bibnamefont {Lipinski~Jusinskas}}, \bibinfo {author} {\bibfnamefont {C.}~\bibnamefont {Lopez-Arcos}},\ and\ \bibinfo {author} {\bibfnamefont {A.}~\bibnamefont {Quintero~Velez}},\ }\bibfield  {title} {\bibinfo {title} {{One-Loop Off-Shell Amplitudes from Classical Equations of Motion}},\ }\href {https://doi.org/10.1103/PhysRevLett.130.081601} {\bibfield  {journal} {\bibinfo  {journal} {Phys. Rev. Lett.}\ }\textbf {\bibinfo {volume} {130}},\ \bibinfo {pages} {081601} (\bibinfo {year} {2023})},\ \Eprint {https://arxiv.org/abs/2208.02831} {arXiv:2208.02831 [hep-th]} \BibitemShut {NoStop}%
\bibitem [{\citenamefont {Gomez}\ \emph {et~al.}(2024)\citenamefont {Gomez}, \citenamefont {Lipinski~Jusinskas}, \citenamefont {Lopez-Arcos},\ and\ \citenamefont {Quintero~Velez}}]{Gomez:2024xec}%
  \BibitemOpen
  \bibfield  {author} {\bibinfo {author} {\bibfnamefont {H.}~\bibnamefont {Gomez}}, \bibinfo {author} {\bibfnamefont {R.}~\bibnamefont {Lipinski~Jusinskas}}, \bibinfo {author} {\bibfnamefont {C.}~\bibnamefont {Lopez-Arcos}},\ and\ \bibinfo {author} {\bibfnamefont {A.}~\bibnamefont {Quintero~Velez}},\ }\bibfield  {title} {\bibinfo {title} {{One-loop $N$-point correlators in pure gravity}},\ }\href@noop {} {\  (\bibinfo {year} {2024})},\ \Eprint {https://arxiv.org/abs/2411.07939} {arXiv:2411.07939 [hep-th]} \BibitemShut {NoStop}%
\bibitem [{\citenamefont {Rosly}\ and\ \citenamefont {Selivanov}(1996)}]{Rosly:1996cp}%
  \BibitemOpen
  \bibfield  {author} {\bibinfo {author} {\bibfnamefont {A.~A.}\ \bibnamefont {Rosly}}\ and\ \bibinfo {author} {\bibfnamefont {K.~G.}\ \bibnamefont {Selivanov}},\ }\bibfield  {title} {\bibinfo {title} {{What we think about multiparticle amplitudes}},\ }\href@noop {} {\  (\bibinfo {year} {1996})},\ \Eprint {https://arxiv.org/abs/hep-th/9610070} {arXiv:hep-th/9610070} \BibitemShut {NoStop}%
\bibitem [{\citenamefont {Rosly}\ and\ \citenamefont {Selivanov}(1997{\natexlab{a}})}]{Rosly:1996vr}%
  \BibitemOpen
  \bibfield  {author} {\bibinfo {author} {\bibfnamefont {A.~A.}\ \bibnamefont {Rosly}}\ and\ \bibinfo {author} {\bibfnamefont {K.~G.}\ \bibnamefont {Selivanov}},\ }\bibfield  {title} {\bibinfo {title} {{On amplitudes in selfdual sector of Yang-Mills theory}},\ }\href {https://doi.org/10.1016/S0370-2693(97)00268-2} {\bibfield  {journal} {\bibinfo  {journal} {Phys. Lett. B}\ }\textbf {\bibinfo {volume} {399}},\ \bibinfo {pages} {135} (\bibinfo {year} {1997}{\natexlab{a}})},\ \Eprint {https://arxiv.org/abs/hep-th/9611101} {arXiv:hep-th/9611101} \BibitemShut {NoStop}%
\bibitem [{\citenamefont {Selivanov}(1998)}]{Selivanov:1997aq}%
  \BibitemOpen
  \bibfield  {author} {\bibinfo {author} {\bibfnamefont {K.~G.}\ \bibnamefont {Selivanov}},\ }\bibfield  {title} {\bibinfo {title} {{SD perturbiner in Yang-Mills + gravity}},\ }\href {https://doi.org/10.1016/S0370-2693(97)01514-1} {\bibfield  {journal} {\bibinfo  {journal} {Phys. Lett. B}\ }\textbf {\bibinfo {volume} {420}},\ \bibinfo {pages} {274} (\bibinfo {year} {1998})},\ \Eprint {https://arxiv.org/abs/hep-th/9710197} {arXiv:hep-th/9710197} \BibitemShut {NoStop}%
\bibitem [{\citenamefont {Rosly}\ and\ \citenamefont {Selivanov}(1997{\natexlab{b}})}]{Rosly:1997ap}%
  \BibitemOpen
  \bibfield  {author} {\bibinfo {author} {\bibfnamefont {A.~A.}\ \bibnamefont {Rosly}}\ and\ \bibinfo {author} {\bibfnamefont {K.~G.}\ \bibnamefont {Selivanov}},\ }\bibfield  {title} {\bibinfo {title} {{Gravitational SD perturbiner}},\ }\href@noop {} {\  (\bibinfo {year} {1997}{\natexlab{b}})},\ \Eprint {https://arxiv.org/abs/hep-th/9710196} {arXiv:hep-th/9710196} \BibitemShut {NoStop}%
\bibitem [{\citenamefont {Mizera}\ and\ \citenamefont {Skrzypek}(2018)}]{Mizera:2018jbh}%
  \BibitemOpen
  \bibfield  {author} {\bibinfo {author} {\bibfnamefont {S.}~\bibnamefont {Mizera}}\ and\ \bibinfo {author} {\bibfnamefont {B.}~\bibnamefont {Skrzypek}},\ }\bibfield  {title} {\bibinfo {title} {{Perturbiner Methods for Effective Field Theories and the Double Copy}},\ }\href {https://doi.org/10.1007/JHEP10(2018)018} {\bibfield  {journal} {\bibinfo  {journal} {JHEP}\ }\textbf {\bibinfo {volume} {10}},\ \bibinfo {pages} {018}},\ \Eprint {https://arxiv.org/abs/1809.02096} {arXiv:1809.02096 [hep-th]} \BibitemShut {NoStop}%
\bibitem [{\citenamefont {Berends}\ and\ \citenamefont {Giele}(1988)}]{Berends:1987me}%
  \BibitemOpen
  \bibfield  {author} {\bibinfo {author} {\bibfnamefont {F.~A.}\ \bibnamefont {Berends}}\ and\ \bibinfo {author} {\bibfnamefont {W.~T.}\ \bibnamefont {Giele}},\ }\bibfield  {title} {\bibinfo {title} {{Recursive Calculations for Processes with n Gluons}},\ }\href {https://doi.org/10.1016/0550-3213(88)90442-7} {\bibfield  {journal} {\bibinfo  {journal} {Nucl. Phys. B}\ }\textbf {\bibinfo {volume} {306}},\ \bibinfo {pages} {759} (\bibinfo {year} {1988})}\BibitemShut {NoStop}%
\bibitem [{\citenamefont {Luna}\ \emph {et~al.}(2017)\citenamefont {Luna}, \citenamefont {Monteiro}, \citenamefont {Nicholson}, \citenamefont {Ochirov}, \citenamefont {O'Connell}, \citenamefont {Westerberg},\ and\ \citenamefont {White}}]{Luna:2016hge}%
  \BibitemOpen
  \bibfield  {author} {\bibinfo {author} {\bibfnamefont {A.}~\bibnamefont {Luna}}, \bibinfo {author} {\bibfnamefont {R.}~\bibnamefont {Monteiro}}, \bibinfo {author} {\bibfnamefont {I.}~\bibnamefont {Nicholson}}, \bibinfo {author} {\bibfnamefont {A.}~\bibnamefont {Ochirov}}, \bibinfo {author} {\bibfnamefont {D.}~\bibnamefont {O'Connell}}, \bibinfo {author} {\bibfnamefont {N.}~\bibnamefont {Westerberg}},\ and\ \bibinfo {author} {\bibfnamefont {C.~D.}\ \bibnamefont {White}},\ }\bibfield  {title} {\bibinfo {title} {{Perturbative spacetimes from Yang-Mills theory}},\ }\href {https://doi.org/10.1007/JHEP04(2017)069} {\bibfield  {journal} {\bibinfo  {journal} {JHEP}\ }\textbf {\bibinfo {volume} {04}},\ \bibinfo {pages} {069}},\ \Eprint {https://arxiv.org/abs/1611.07508} {arXiv:1611.07508 [hep-th]} \BibitemShut {NoStop}%
\bibitem [{\citenamefont {Monteiro}\ and\ \citenamefont {O'Connell}(2011)}]{Monteiro:2011pc}%
  \BibitemOpen
  \bibfield  {author} {\bibinfo {author} {\bibfnamefont {R.}~\bibnamefont {Monteiro}}\ and\ \bibinfo {author} {\bibfnamefont {D.}~\bibnamefont {O'Connell}},\ }\bibfield  {title} {\bibinfo {title} {{The Kinematic Algebra From the Self-Dual Sector}},\ }\href {https://doi.org/10.1007/JHEP07(2011)007} {\bibfield  {journal} {\bibinfo  {journal} {JHEP}\ }\textbf {\bibinfo {volume} {07}},\ \bibinfo {pages} {007}},\ \Eprint {https://arxiv.org/abs/1105.2565} {arXiv:1105.2565 [hep-th]} \BibitemShut {NoStop}%
\bibitem [{\citenamefont {Borsten}\ \emph {et~al.}(2023)\citenamefont {Borsten}, \citenamefont {Jurco}, \citenamefont {Kim}, \citenamefont {Macrelli}, \citenamefont {Saemann},\ and\ \citenamefont {Wolf}}]{Borsten:2023paw}%
  \BibitemOpen
  \bibfield  {author} {\bibinfo {author} {\bibfnamefont {L.}~\bibnamefont {Borsten}}, \bibinfo {author} {\bibfnamefont {B.}~\bibnamefont {Jurco}}, \bibinfo {author} {\bibfnamefont {H.}~\bibnamefont {Kim}}, \bibinfo {author} {\bibfnamefont {T.}~\bibnamefont {Macrelli}}, \bibinfo {author} {\bibfnamefont {C.}~\bibnamefont {Saemann}},\ and\ \bibinfo {author} {\bibfnamefont {M.}~\bibnamefont {Wolf}},\ }\bibfield  {title} {\bibinfo {title} {{Double-copying self-dual Yang-Mills theory to self-dual gravity on twistor space}},\ }\href {https://doi.org/10.1007/JHEP11(2023)172} {\bibfield  {journal} {\bibinfo  {journal} {JHEP}\ }\textbf {\bibinfo {volume} {11}},\ \bibinfo {pages} {172}},\ \Eprint {https://arxiv.org/abs/2307.10383} {arXiv:2307.10383 [hep-th]} \BibitemShut {NoStop}%
\bibitem [{\citenamefont {Frost}\ \emph {et~al.}(2023)\citenamefont {Frost}, \citenamefont {Mafra},\ and\ \citenamefont {Mason}}]{Frost:2020eoa}%
  \BibitemOpen
  \bibfield  {author} {\bibinfo {author} {\bibfnamefont {H.}~\bibnamefont {Frost}}, \bibinfo {author} {\bibfnamefont {C.~R.}\ \bibnamefont {Mafra}},\ and\ \bibinfo {author} {\bibfnamefont {L.}~\bibnamefont {Mason}},\ }\bibfield  {title} {\bibinfo {title} {{A Lie Bracket for the Momentum Kernel}},\ }\href {https://doi.org/10.1007/s00220-023-04748-z} {\bibfield  {journal} {\bibinfo  {journal} {Commun. Math. Phys.}\ }\textbf {\bibinfo {volume} {402}},\ \bibinfo {pages} {1307} (\bibinfo {year} {2023})},\ \Eprint {https://arxiv.org/abs/2012.00519} {arXiv:2012.00519 [hep-th]} \BibitemShut {NoStop}%
\bibitem [{\citenamefont {Plebanski}(1975)}]{Plebanski:1975wn}%
  \BibitemOpen
  \bibfield  {author} {\bibinfo {author} {\bibfnamefont {J.~F.}\ \bibnamefont {Plebanski}},\ }\bibfield  {title} {\bibinfo {title} {{Some solutions of complex Einstein equations}},\ }\href {https://doi.org/10.1063/1.522505} {\bibfield  {journal} {\bibinfo  {journal} {J. Math. Phys.}\ }\textbf {\bibinfo {volume} {16}},\ \bibinfo {pages} {2395} (\bibinfo {year} {1975})}\BibitemShut {NoStop}%
\bibitem [{\citenamefont {Bardeen}(1996)}]{Bardeen:1995gk}%
  \BibitemOpen
  \bibfield  {author} {\bibinfo {author} {\bibfnamefont {W.~A.}\ \bibnamefont {Bardeen}},\ }\bibfield  {title} {\bibinfo {title} {{Selfdual Yang-Mills theory, integrability and multiparton amplitudes}},\ }\href {https://doi.org/10.1143/PTPS.123.1} {\bibfield  {journal} {\bibinfo  {journal} {Prog. Theor. Phys. Suppl.}\ }\textbf {\bibinfo {volume} {123}},\ \bibinfo {pages} {1} (\bibinfo {year} {1996})}\BibitemShut {NoStop}%
\bibitem [{\citenamefont {Cangemi}(1997)}]{Cangemi:1996rx}%
  \BibitemOpen
  \bibfield  {author} {\bibinfo {author} {\bibfnamefont {D.}~\bibnamefont {Cangemi}},\ }\bibfield  {title} {\bibinfo {title} {{Selfdual Yang-Mills theory and one loop like - helicity QCD multi - gluon amplitudes}},\ }\href {https://doi.org/10.1016/S0550-3213(96)00586-X} {\bibfield  {journal} {\bibinfo  {journal} {Nucl. Phys. B}\ }\textbf {\bibinfo {volume} {484}},\ \bibinfo {pages} {521} (\bibinfo {year} {1997})},\ \Eprint {https://arxiv.org/abs/hep-th/9605208} {arXiv:hep-th/9605208} \BibitemShut {NoStop}%
\bibitem [{\citenamefont {Cho}\ \emph {et~al.}(2022)\citenamefont {Cho}, \citenamefont {Kim},\ and\ \citenamefont {Lee}}]{Cho:2021nim}%
  \BibitemOpen
  \bibfield  {author} {\bibinfo {author} {\bibfnamefont {K.}~\bibnamefont {Cho}}, \bibinfo {author} {\bibfnamefont {K.}~\bibnamefont {Kim}},\ and\ \bibinfo {author} {\bibfnamefont {K.}~\bibnamefont {Lee}},\ }\bibfield  {title} {\bibinfo {title} {{The off-shell recursion for gravity and the classical double copy for currents}},\ }\href {https://doi.org/10.1007/JHEP01(2022)186} {\bibfield  {journal} {\bibinfo  {journal} {JHEP}\ }\textbf {\bibinfo {volume} {01}},\ \bibinfo {pages} {186}},\ \Eprint {https://arxiv.org/abs/2109.06392} {arXiv:2109.06392 [hep-th]} \BibitemShut {NoStop}%
\bibitem [{\citenamefont {Correa}\ \emph {et~al.}(2024)\citenamefont {Correa}, \citenamefont {Lopez-Arcos},\ and\ \citenamefont {Quintero~Velez}}]{Correa:2024mub}%
  \BibitemOpen
  \bibfield  {author} {\bibinfo {author} {\bibfnamefont {D.~H.}\ \bibnamefont {Correa}}, \bibinfo {author} {\bibfnamefont {C.}~\bibnamefont {Lopez-Arcos}},\ and\ \bibinfo {author} {\bibfnamefont {A.}~\bibnamefont {Quintero~Velez}},\ }\bibfield  {title} {\bibinfo {title} {{Tree- and one-loop-level double copy for the (anti)self-dual sectors of Yang-Mills and gravity}},\ }\href@noop {} {\  (\bibinfo {year} {2024})},\ \Eprint {https://arxiv.org/abs/2412.07498} {arXiv:2412.07498 [hep-th]} \BibitemShut {NoStop}%
\bibitem [{\citenamefont {Krasnov}(2017)}]{Krasnov:2016emc}%
  \BibitemOpen
  \bibfield  {author} {\bibinfo {author} {\bibfnamefont {K.}~\bibnamefont {Krasnov}},\ }\bibfield  {title} {\bibinfo {title} {{Self-Dual Gravity}},\ }\href {https://doi.org/10.1088/1361-6382/aa65e5} {\bibfield  {journal} {\bibinfo  {journal} {Class. Quant. Grav.}\ }\textbf {\bibinfo {volume} {34}},\ \bibinfo {pages} {095001} (\bibinfo {year} {2017})},\ \Eprint {https://arxiv.org/abs/1610.01457} {arXiv:1610.01457 [hep-th]} \BibitemShut {NoStop}%
\bibitem [{\citenamefont {Bakas}(1989)}]{Bakas:1989xu}%
  \BibitemOpen
  \bibfield  {author} {\bibinfo {author} {\bibfnamefont {I.}~\bibnamefont {Bakas}},\ }\bibfield  {title} {\bibinfo {title} {{The Large n Limit of Extended Conformal Symmetries}},\ }\href {https://doi.org/10.1016/0370-2693(89)90525-X} {\bibfield  {journal} {\bibinfo  {journal} {Phys. Lett. B}\ }\textbf {\bibinfo {volume} {228}},\ \bibinfo {pages} {57} (\bibinfo {year} {1989})}\BibitemShut {NoStop}%
\bibitem [{\citenamefont {Berends}\ and\ \citenamefont {Giele}(1989)}]{Berends:1988zn}%
  \BibitemOpen
  \bibfield  {author} {\bibinfo {author} {\bibfnamefont {F.~A.}\ \bibnamefont {Berends}}\ and\ \bibinfo {author} {\bibfnamefont {W.~T.}\ \bibnamefont {Giele}},\ }\bibfield  {title} {\bibinfo {title} {{Multiple Soft Gluon Radiation in Parton Processes}},\ }\href {https://doi.org/10.1016/0550-3213(89)90398-2} {\bibfield  {journal} {\bibinfo  {journal} {Nucl. Phys. B}\ }\textbf {\bibinfo {volume} {313}},\ \bibinfo {pages} {595} (\bibinfo {year} {1989})}\BibitemShut {NoStop}%
\bibitem [{\citenamefont {Lee}\ \emph {et~al.}(2016)\citenamefont {Lee}, \citenamefont {Mafra},\ and\ \citenamefont {Schlotterer}}]{Lee:2015upy}%
  \BibitemOpen
  \bibfield  {author} {\bibinfo {author} {\bibfnamefont {S.}~\bibnamefont {Lee}}, \bibinfo {author} {\bibfnamefont {C.~R.}\ \bibnamefont {Mafra}},\ and\ \bibinfo {author} {\bibfnamefont {O.}~\bibnamefont {Schlotterer}},\ }\bibfield  {title} {\bibinfo {title} {{Non-linear gauge transformations in $D=10$ SYM theory and the BCJ duality}},\ }\href {https://doi.org/10.1007/JHEP03(2016)090} {\bibfield  {journal} {\bibinfo  {journal} {JHEP}\ }\textbf {\bibinfo {volume} {03}},\ \bibinfo {pages} {090}},\ \Eprint {https://arxiv.org/abs/1510.08843} {arXiv:1510.08843 [hep-th]} \BibitemShut {NoStop}%
\bibitem [{\citenamefont {Mafra}\ and\ \citenamefont {Schlotterer}(2016)}]{Mafra:2015vca}%
  \BibitemOpen
  \bibfield  {author} {\bibinfo {author} {\bibfnamefont {C.~R.}\ \bibnamefont {Mafra}}\ and\ \bibinfo {author} {\bibfnamefont {O.}~\bibnamefont {Schlotterer}},\ }\bibfield  {title} {\bibinfo {title} {{Berends-Giele recursions and the BCJ duality in superspace and components}},\ }\href {https://doi.org/10.1007/JHEP03(2016)097} {\bibfield  {journal} {\bibinfo  {journal} {JHEP}\ }\textbf {\bibinfo {volume} {03}},\ \bibinfo {pages} {097}},\ \Eprint {https://arxiv.org/abs/1510.08846} {arXiv:1510.08846 [hep-th]} \BibitemShut {NoStop}%
\bibitem [{\citenamefont {Mafra}(2020)}]{Mafra:2020qst}%
  \BibitemOpen
  \bibfield  {author} {\bibinfo {author} {\bibfnamefont {C.~R.}\ \bibnamefont {Mafra}},\ }\bibfield  {title} {\bibinfo {title} {{Planar binary trees in scattering amplitudes}}\ }\href {https://doi.org/10.4171/205-1/6} {10.4171/205-1/6} (\bibinfo {year} {2020}),\ \Eprint {https://arxiv.org/abs/2011.14413} {arXiv:2011.14413 [math.CO]} \BibitemShut {NoStop}%
\bibitem [{\citenamefont {Escudero}\ \emph {et~al.}(2023)\citenamefont {Escudero}, \citenamefont {Lopez-Arcos},\ and\ \citenamefont {Quintero~Velez}}]{Escudero:2022zdz}%
  \BibitemOpen
  \bibfield  {author} {\bibinfo {author} {\bibfnamefont {V.~G.}\ \bibnamefont {Escudero}}, \bibinfo {author} {\bibfnamefont {C.}~\bibnamefont {Lopez-Arcos}},\ and\ \bibinfo {author} {\bibfnamefont {A.}~\bibnamefont {Quintero~Velez}},\ }\bibfield  {title} {\bibinfo {title} {{Homotopy double copy and the Kawai\textendash{}Lewellen\textendash{}Tye relations for the non-abelian and tensor Navier\textendash{}Stokes equations}},\ }\href {https://doi.org/10.1063/5.0119508} {\bibfield  {journal} {\bibinfo  {journal} {J. Math. Phys.}\ }\textbf {\bibinfo {volume} {64}},\ \bibinfo {pages} {032304} (\bibinfo {year} {2023})},\ \Eprint {https://arxiv.org/abs/2201.06047} {arXiv:2201.06047 [math-ph]} \BibitemShut {NoStop}%
\bibitem [{\citenamefont {Bern}\ \emph {et~al.}(1999)\citenamefont {Bern}, \citenamefont {Dixon}, \citenamefont {Perelstein},\ and\ \citenamefont {Rozowsky}}]{Bern:1998sv}%
  \BibitemOpen
  \bibfield  {author} {\bibinfo {author} {\bibfnamefont {Z.}~\bibnamefont {Bern}}, \bibinfo {author} {\bibfnamefont {L.~J.}\ \bibnamefont {Dixon}}, \bibinfo {author} {\bibfnamefont {M.}~\bibnamefont {Perelstein}},\ and\ \bibinfo {author} {\bibfnamefont {J.~S.}\ \bibnamefont {Rozowsky}},\ }\bibfield  {title} {\bibinfo {title} {{Multileg one loop gravity amplitudes from gauge theory}},\ }\href {https://doi.org/10.1016/S0550-3213(99)00029-2} {\bibfield  {journal} {\bibinfo  {journal} {Nucl. Phys. B}\ }\textbf {\bibinfo {volume} {546}},\ \bibinfo {pages} {423} (\bibinfo {year} {1999})},\ \Eprint {https://arxiv.org/abs/hep-th/9811140} {arXiv:hep-th/9811140} \BibitemShut {NoStop}%
\bibitem [{\citenamefont {Bjerrum-Bohr}\ \emph {et~al.}(2010)\citenamefont {Bjerrum-Bohr}, \citenamefont {Damgaard}, \citenamefont {Feng},\ and\ \citenamefont {Sondergaard}}]{Bjerrum-Bohr:2010diw}%
  \BibitemOpen
  \bibfield  {author} {\bibinfo {author} {\bibfnamefont {N.~E.~J.}\ \bibnamefont {Bjerrum-Bohr}}, \bibinfo {author} {\bibfnamefont {P.~H.}\ \bibnamefont {Damgaard}}, \bibinfo {author} {\bibfnamefont {B.}~\bibnamefont {Feng}},\ and\ \bibinfo {author} {\bibfnamefont {T.}~\bibnamefont {Sondergaard}},\ }\bibfield  {title} {\bibinfo {title} {{Gravity and Yang-Mills Amplitude Relations}},\ }\href {https://doi.org/10.1103/PhysRevD.82.107702} {\bibfield  {journal} {\bibinfo  {journal} {Phys. Rev. D}\ }\textbf {\bibinfo {volume} {82}},\ \bibinfo {pages} {107702} (\bibinfo {year} {2010})},\ \Eprint {https://arxiv.org/abs/1005.4367} {arXiv:1005.4367 [hep-th]} \BibitemShut {NoStop}%
\bibitem [{\citenamefont {Gomez}\ and\ \citenamefont {Jusinskas}(2021)}]{Gomez:2021shh}%
  \BibitemOpen
  \bibfield  {author} {\bibinfo {author} {\bibfnamefont {H.}~\bibnamefont {Gomez}}\ and\ \bibinfo {author} {\bibfnamefont {R.~L.}\ \bibnamefont {Jusinskas}},\ }\bibfield  {title} {\bibinfo {title} {{Multiparticle Solutions to Einstein\textquoteright{}s Equations}},\ }\href {https://doi.org/10.1103/PhysRevLett.127.181603} {\bibfield  {journal} {\bibinfo  {journal} {Phys. Rev. Lett.}\ }\textbf {\bibinfo {volume} {127}},\ \bibinfo {pages} {181603} (\bibinfo {year} {2021})},\ \Eprint {https://arxiv.org/abs/2106.12584} {arXiv:2106.12584 [hep-th]} \BibitemShut {NoStop}%
\bibitem [{\citenamefont {Miller}(2024)}]{Miller:2024oza}%
  \BibitemOpen
  \bibfield  {author} {\bibinfo {author} {\bibfnamefont {N.}~\bibnamefont {Miller}},\ }\bibfield  {title} {\bibinfo {title} {{Proof of the graviton MHV formula using Plebanski's second heavenly equation}},\ }\href@noop {} {\  (\bibinfo {year} {2024})},\ \Eprint {https://arxiv.org/abs/2408.11139} {arXiv:2408.11139 [hep-th]} \BibitemShut {NoStop}%
\bibitem [{\citenamefont {Berman}\ \emph {et~al.}(2019)\citenamefont {Berman}, \citenamefont {Chac\'on}, \citenamefont {Luna},\ and\ \citenamefont {White}}]{Berman:2018hwd}%
  \BibitemOpen
  \bibfield  {author} {\bibinfo {author} {\bibfnamefont {D.~S.}\ \bibnamefont {Berman}}, \bibinfo {author} {\bibfnamefont {E.}~\bibnamefont {Chac\'on}}, \bibinfo {author} {\bibfnamefont {A.}~\bibnamefont {Luna}},\ and\ \bibinfo {author} {\bibfnamefont {C.~D.}\ \bibnamefont {White}},\ }\bibfield  {title} {\bibinfo {title} {{The self-dual classical double copy, and the Eguchi-Hanson instanton}},\ }\href {https://doi.org/10.1007/JHEP01(2019)107} {\bibfield  {journal} {\bibinfo  {journal} {JHEP}\ }\textbf {\bibinfo {volume} {01}},\ \bibinfo {pages} {107}},\ \Eprint {https://arxiv.org/abs/1809.04063} {arXiv:1809.04063 [hep-th]} \BibitemShut {NoStop}%
\bibitem [{\citenamefont {Penrose}(1976)}]{Penrose:1976js}%
  \BibitemOpen
  \bibfield  {author} {\bibinfo {author} {\bibfnamefont {R.}~\bibnamefont {Penrose}},\ }\bibfield  {title} {\bibinfo {title} {{Nonlinear Gravitons and Curved Twistor Theory}},\ }\href {https://doi.org/10.1007/BF00762011} {\bibfield  {journal} {\bibinfo  {journal} {Gen. Rel. Grav.}\ }\textbf {\bibinfo {volume} {7}},\ \bibinfo {pages} {31} (\bibinfo {year} {1976})}\BibitemShut {NoStop}%
\end{thebibliography}%

\end{document}